\documentclass[12pt]{iopart}

\pdfoutput=1

\usepackage{graphicx,sidecap}
\usepackage{bm}
\usepackage{braket}
\usepackage{amssymb}
\usepackage{amsfonts}
\usepackage{mathrsfs}
\usepackage{xcolor}
\usepackage{hyperref}
\usepackage{enumerate}
\usepackage[toc,title]{appendix}

\catcode`@=11 
\renewcommand\tableofcontents{%
  \section*{\contentsname}%
  \@starttoc{toc}%
}
\catcode`@=12

\def\be{\begin{equation}}
\def\ee{\end{equation}}

\def\bea{\begin{eqnarray}}
\def\eea{\end{eqnarray}}

\def\Tr{{\rm Tr}}

\usepackage{xparse}

\DeclareDocumentCommand{\TrProd}{ m O{} o O{} o O{} o O{} o }{%
\{ \Gamma_{#1}^{#2}
\IfNoValueTF{#3}{}{,\Gamma_{#3}^{#4}}
\IfNoValueTF{#5}{}{,\Gamma_{#5}^{#6}}
\IfNoValueTF{#7}{}{,\Gamma_{#7}^{#8}}
\IfNoValueTF{#9}{}{,\Gamma_{#9}}
\}
}
\DeclareDocumentCommand{\TrProdNorm}{ m }{%
\{ \Gamma_{1}^{#1} \}
}

\DeclareDocumentCommand{\TrProdTilde}{ m O{} o O{} o O{} o O{} o }{%
\{ \widetilde\Gamma_{#1}^{#2}
\IfNoValueTF{#3}{}{,\widetilde\Gamma_{#3}^{#4}}
\IfNoValueTF{#5}{}{,\widetilde\Gamma_{#5}^{#6}}
\IfNoValueTF{#7}{}{,\widetilde\Gamma_{#7}^{#8}}
\IfNoValueTF{#9}{}{,\widetilde\Gamma_{#9}}
\}
}
\DeclareDocumentCommand{\TrProdNormTilde}{ m }{%
\{ \widetilde\Gamma_{1}^{#1} \}
}

\DeclareDocumentCommand{\ch}{ m o o o m o o o }{%
\begin{bmatrix}
#1 \IfNoValueTF{#2}{}{& #2}\IfNoValueTF{#3}{}{& #3}\IfNoValueTF{#4}{}{& #4} \\
#5 \IfNoValueTF{#6}{}{& #6}\IfNoValueTF{#7}{}{& #7}\IfNoValueTF{#8}{}{& #8}
\end{bmatrix}_{\tau}
}

\begin{document}

\title[
Partial transpose of two disjoint blocks in XY spin chains
]{
\\
Partial transpose of two disjoint blocks  in XY spin chains
}

\vspace{.5cm}

\author{Andrea Coser, Erik Tonni and Pasquale Calabrese}
\address{SISSA and INFN, via Bonomea 265, 34136 Trieste, Italy. }

\vspace{.5cm}

\begin{abstract}
We consider the partial transpose of the spin reduced density matrix of two disjoint blocks in spin chains admitting a 
representation in terms of free fermions, such as XY chains. 
We exploit the solution of the model in terms of Majorana fermions and show that such partial transpose in the spin variables 
is a linear combination of four Gaussian fermionic  operators. 
This representation allows to explicitly construct and evaluate the integer moments of the partial transpose. 
We numerically study critical XX and Ising chains and we show that the asymptotic results for large blocks agree with 
conformal field theory predictions if corrections to the scaling are properly taken into account. 

\end{abstract}

\maketitle


\section{Introduction}

In the last decade, the understanding of the entanglement content of extended quantum systems 
boosted an intense research activity at the boundary between condensed matter, quantum information and 
quantum field theory (see e.g. Refs. \cite{rev} as reviews).
The bipartite entanglement for an extended system in a pure state is measured by the well-known entanglement 
entropy which is the von Neumann entropy corresponding to the reduced density matrix of one of the two parts. 
One of the most remarkable results in this field is the logarithmic divergence of the entanglement with the 
subsystem size in the case when the low energy properties of the extended critical quantum systems are described by 
a 1+1 dimensional conformal invariant theory \cite{Holzhey,vlrk-03,cc-04,cc-rev}. 

Conversely, when an extended quantum system is in a mixed state (or one considers a tripartition of a pure state 
and is interested in the relative entanglement between two of the three parts) the quantification of the entanglement 
is much more complicated. 
A very useful concept is that of partial transposition. 
Indeed it has been shown that the presence of entanglement in a bipartite mixed state 
is related to occurrence of negative eigenvalues in the spectrum of the partial transpose of the density matrix \cite{partial}. 
This led to the proposal of the negativity \cite{vw-02} (or the logarithmic negativity) which was later shown to be 
an entanglement monotone \cite{neg-mon}, i.e. a good entanglement measure from a quantum information perspective. 
Compared to other entanglement measurements for mixed states, the negativity has the important property of being
easily calculable for an arbitrary quantum state once its density matrix is known (and indeed for this reason it 
has been named a ``computable measure of entanglement" \cite{vw-02}). 

Recently, a systematic path integral approach to construct the partial transpose of the reduced density matrix 
has been developed and from this the negativity in 1+1 dimensional relativistic quantum field theories 
is obtained via a replica trick \cite{cct-neg-letter}.  
This approach has been successfully applied to the study of one-dimensional conformal field theories (CFT) 
in the ground state \cite{cct-neg-letter,cct-neg-long}, in thermal state \cite{cct-neg-T,ez-14}, and in 
non-equilibrium protocols \cite{ez-14,ctc-14,hd-15,wcr-15}, as well as to topological systems \cite{c-13,lv-13}.
The CFT predictions have been tested for several models 
\cite{cct-neg-long,cct-neg-T,ctc-14,ctt-13,a-13,AlbaLauchlin-neg},
especially against exact results \cite{cct-neg-long,cct-neg-T,ctc-14,mrpr-09} for free bosonic 
systems (such as the harmonic chain).
Indeed for free bosonic models, the partial transposition corresponds to a time-reversal operation  leading to a partially 
transposed reduced density matrix which is Gaussian \cite{aepw-02} and that can be straightforwardly diagonalised by 
correlation matrix techniques \cite{pc-99,br-04,pe-09}.
It should be also mentioned that there exist some earlier results for the negativity in many body systems 
\cite{mrpr-09,aepw-02,br-04,wmb-09,aw-08,fcga-08,Neg3,sod,kor}.

In the case of free fermionic systems (such as the tight-binding model and XY spin chains)
the calculation of the negativity is instead much more involved. 
Indeed the partial  transpose of the reduced density matrix is not a Gaussian operator and standard techniques 
based on the correlation matrix \cite{pe-09} cannot be applied. 
In view of the importance that exact calculations for free fermionic systems played in the understanding 
of the entanglement entropy \cite{vlrk-03,jk-04,ij-08,gl-rev,ce-10,cmv-11}, 
it is highly desirable to have an exact representation of the negativity 
also for free fermionic systems. 
A major step in this direction has been very recently achieved by Eisler and Zimboras \cite{ez-15}
who showed that the partial transpose is a linear combination of two Gaussian operators. 
Unfortunately, it is still not possible to extract the spectrum of the partial transpose and hence the negativity,
but at least one can access to integer powers of the partial transpose which are the main 
ingredient for the replica approach to negativity. 

In Ref. \cite{ez-15} only truly fermionic systems have been considered and not spin chains that can be 
mapped to a fermionic system by means of a (non-local) Jordan-Wigner transformation. 
Indeed, in the very interesting case of two disjoint blocks in a spin chain the density matrix of 
spins and fermions are not equal \cite{atc-10,ip-10,fc-10} and this consequently affects also the partial transposition, as 
already pointed out in \cite{ez-15}. 
In this manuscript we fill this gap by giving an exact representation of the partial transpose of the 
reduced density matrix for two disjoint blocks in the XY spin chain and from this we calculate the traces of its
integer powers. These turn out to converge to the CFT predictions in the limit of large intervals.  

The manuscript is organised as follows. 
In Sec. \ref{sec2} we describe the model and the definition of the quantities we will study. 
In Sec. \ref{sec renyi} we review the results of Ref. \cite{fc-10} for the moments of the spin reduced density matrix 
of two disjoint blocks. 
In Sec. \ref{sec pt2} we move to the core of this manuscript deriving an explicit representation of the partial transpose 
of the spin reduced density matrix as a sum of four Gaussian fermionic matrices. 
This allows to obtain explicit representations for the moments  of the partial transpose.
In Sec. \ref{sec num} we use the above results to numerically calculate these moments up to $n=5$ for the 
critical Ising model and XX chain and carefully compare them with CFT predictions by taking into account 
corrections to the scaling. 
In Sec. \ref{sec FF} we numerically evaluate and study the moments of the partial transpose for two disjoint blocks 
of fermions and again we compare with new CFT  predictions. 
Finally in Sec. \ref{concl} we draw our conclusions. 
In appendix \ref{app1} we report all the CFT results which we needed in this manuscript.

\section{The model and the quantities of interest}
\label{sec2}

In this manuscript we consider  the XY spin chains with Hamiltonian
\be
\label{H_XY}
H_{XY} = -\frac{1}{2} 
\sum_{j=1}^L \left(\,
\frac{1+\gamma}{2}\, \sigma_j^x \sigma_{j+1}^x 
+ \frac{1-\gamma}{2}\, \sigma_j^y \sigma_{j+1}^y
+h \,\sigma_j^z
\right),
\ee
where $\sigma_{j}^\alpha$ are the Pauli matrices at the $j$-th site and we assume periodic boundary conditions 
$\sigma_{L+1}^\alpha=\sigma_1^\alpha $.
For $\gamma=1$ Eq. (\ref{H_XY}) reduces to the Hamiltonian of the Ising model in a transverse field while 
for $\gamma=0$ to the one of the XX spin chain.  
The Hamiltonian (\ref{H_XY}) is a paradigmatic model for quantum phase transitions \cite{sach-book}. 
In fact, it  depends on two parameters: the transverse magnetic field $h$ and the anisotropy parameter $\gamma$. 
The system is critical for $h=1$ and any $\gamma$ with a transition that belongs to the Ising universality class.
It is also critical for $\gamma=0$ and $|h|<1$ with a continuum limit given by a free compactified boson. 

The Jordan-Wigner transformation 
\be
\label{JW map}
c_j = 
\Big( \prod_{m<j} \sigma_m^z \Big) 
\frac{\sigma_j^x - \textrm{i} \sigma_j^z}{2},
\qquad
c_j^\dagger = 
\Big( \prod_{m<j} \sigma_m^z \Big) 
\frac{\sigma_j^x + \textrm{i} \sigma_j^z}{2},
\ee
maps the spin variables into anti-commuting fermionic ones $\{c_i, c^\dagger_j\}=\delta_{ij}$.
In terms of these fermionic variables the Hamiltonian (\ref{H_XY}) becomes 
\be
H_{XY}=\sum_{i=1}^L \left(\frac12 \left[\gamma c^\dagger_i c^\dagger_{i+1}+ \gamma c_{i+1} c_{i}
+c^\dagger_i c_{i+1}+ c^\dagger_{i+1} c_{i} \right]-h c^\dagger_i c_i
\right),
\label{HXYF}
\ee
where we neglected boundary and additive terms.
This Hamiltonian is quadratic in the fermionic operators 
and hence can be straightforwardly diagonalised in momentum space by means of a Bogoliubov transformation. 

For the study of the reduced density matrices it is very useful to introduce 
the Majorana fermions \cite{vlrk-03}
\be
a_{2j} = c_j + c_j^\dagger,
\qquad
a_{2j-1} = \textrm{i} (c_j - c_j^\dagger),
\ee
which satisfy the anti-commutation relations $\{a_r, a_s\}=2\delta_{rs}$.

\subsection{Quantities of interest}

\begin{figure}
\includegraphics[width=0.99\textwidth]{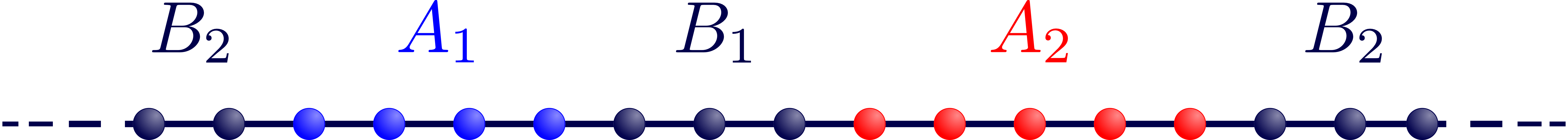}
\caption{We consider the entanglement between two disjoint spin blocks $A_1$ and $A_2$ embedded in a spin chain of arbitrary length.
The reminder of the system is denoted by $B$ which is also composed of two disconnected pieces $B_1$ and $B_2$.
} 
\label{fig:sc}
\end{figure}

The main goal of this manuscript is to determine the entanglement between two disjoint intervals in the XY spin chain.
We consider the geometry depicted in Fig. \ref{fig:sc}: a spin chain is divided in two parts 
$A$ and $B$ and each of them is composed of disconnected pieces.
We denote by $A_1$ and $A_2$ the two blocks in $A=A_1\cup A_2$, $B_1$ is the block in $B$ separating them, 
while $B_2$ is the remainder. 

The reduced density matrix of $A$ is $\rho_A={\rm Tr}_B \rho={\rm Tr}_B |\Psi\rangle\langle \Psi|$, where we are mainly 
interested in the case
in which $|\Psi \rangle$ is the ground state of the XY chain, even if the results of this paper apply to more general 
cases such as excited states, non-equilibrium configurations, finite temperature etc. 
The bipartite entanglement between $A$ and $B$ is given by the well-known entanglement entropy
\be
S_A=-{\rm Tr} \rho_A \ln \rho_A\,,
\ee
or equivalently by the R\'enyi entropies 
\be
S_A^{(n)}=\frac1{1-n}\ln{\rm Tr}\,\rho_A^n \,,
\label{Sndef}
\ee
which in the limit $n\to1$ reduce to the entanglement entropy, but provide more information since it is related to the 
full spectrum of $\rho_A$ \cite{cl-08}.

In the case of two disjoint blocks in the XY spin chain, the R\'enyi entropies for integer $n$ (or equivalently the moments of the 
reduced density matrix $\rho_A$) have been explicitly constructed in Ref. \cite{fc-10}.
However it is still not possible to find the analytic continuation to arbitrary complex values of $n$ and consequently the 
entanglement entropy. This is very similar to the CFT counterpart where also one can calculate only integer moments of the reduced 
density matrices (see Appendix \ref{app1} for a summary of the CFT results of interest for this paper).

However, we are here interested in the entanglement between $A_1$ and $A_2$. 
A measure of this entanglement is provided by the logarithmic negativity defined as follows.
Let us denote by $|e_i^{(1)}\rangle$ and $|e_j^{(2)}\rangle$ two arbitrary bases 
in the Hilbert spaces corresponding to $A_1$ and $A_2$. 
The partial transpose of  $\rho_A$ with respect to $A_2$ degrees of freedom is defined as
\be 
\langle e_i^{(1)} e_j^{(2)}|\rho^{T_2}_A|e_k^{(1)} e_l^{(2)}\rangle=\langle e_i^{(1)} e_l^{(2)}|\rho_A| e^{(1)}_k e^{(2)}_j\rangle,
\ee
and then the {\it logarithmic} negativity as
\be
{\cal E}\equiv\ln ||\rho^{T_2}_A||=\ln \Tr |\rho^{T_2}_A|\,,
\ee
where the trace norm  $||\rho^{T_2}_A||$ is
the sum of the absolute values of the eigenvalues of $\rho^{T_2}_A$.

A systematic method to compute the negativity in quantum field theories has been developed in Ref. \cite{cct-neg-letter, cct-neg-long}
and it is again based on a replica trick from the integer moments of the partial transpose 
\be
\Tr (\rho^{T_2}_A)^n,
\ee 
which turn out to admit different analytic continuations from even and odd $n$ (usually denoted as $n_e$ and $n_o$ respectively).
Consequently, the logarithmic negativity is given by the replica limit
\be
\mathcal{E} = \lim_{n_e \to 1} \ln \Tr \big(\rho_A^{T_2}\big)^{n_e} ,
\ee
performed on the even sequence of moments. 
Unfortunately, in the case of two disjoint intervals, it is very difficult to perform this analytic 
continuation (see Appendix \ref{app1}), although the integer moments $\Tr (\rho_A^{T_2})^n$ are known for the most relevant 
conformal field theories. 
It is however possible to extract very useful information about entanglement and about the partial transpose already from 
the knowledge of the moments~\cite{cct-neg-letter, cct-neg-long}.

Finally, in Refs. \cite{cct-neg-letter, cct-neg-long} the ratios 
\be
\label{R_n def}
R_n 
\equiv \frac{\Tr (\rho^{T_2}_A)^n }{\Tr \rho_A^n },
\ee
have been introduced because of some cancellations (see Appendix \ref{app1}) and since anyhow
\be
\mathcal{E} = \lim_{n_e \to 1} \ln(R_{n_e}) ,
\ee
given that $\Tr \rho_A=1$.


\section{R\'enyi entropies}
\label{sec renyi}

In this section we review the results of Ref. \cite{fc-10}, where the R\'enyi entropies of two disjoint blocks 
$A=A_1\cup A_2$ for the XY spin chains have been computed. 
These results are the main ingredients to construct the integer powers of the partial transpose which will be 
derived in the following section.  
We will denote the number of spins in $A_1$ and $A_2$ with $\ell_1$ and $\ell_2$ respectively and 
the remainder of the system $B$ contains a region separating $A_1$ and $A_2$ denoted as $B_1$, as 
pictorially depicted in Fig. \ref{fig:sc}.

In a general spin $1/2$ chain the 
reduced density matrix $\rho_A =\Tr_B |\Psi \rangle \langle \Psi |$ of 
$A=A_1\cup A_2$ can be computed by summing all the operators in $A$ as follows \cite{vlrk-03}
\be
\label{rhoA correlators}
\rho_A = \frac{1}{2^{\ell_1+\ell_2}} \sum_{\nu_j} 
\Big\langle \prod_{j \in A} \sigma_j^{\nu_j} \Big\rangle 
 \prod_{j \in A} \sigma_j^{\nu_j} ,
\ee
where $j$ is the index labelling the lattice sites and $\nu_j \in \{ 0,1,2,3 \}$, with $\sigma^{0} = {\bf 1}$  
the identity matrix and $\sigma^{1} = \sigma^{x}$, $\sigma^{2} = \sigma^{y}$ and $\sigma^{3} = \sigma^{z}$ the Pauli matrices. 
The multipoints correlators in (\ref{rhoA correlators}) are very difficult to compute, unless there is a representation of the state 
in terms of free fermions. 

For the single interval case, the Jordan-Wigner string $\prod_{m<j} \sigma_m^z $ in Eq. (\ref{JW map}) maps the first 
$\ell$ spins into the first $\ell$ fermions \cite{vlrk-03} so the spin and fermionic entropy are the same.
In general,  calculating a fermionic density matrix is very easy. 
Indeed by Wick theorem they assume a Gaussian form: 
\be
\rho_W = \frac{\exp \big( \frac{1}{4} \sum_{r,s} a_r W_{rs} a_s\big)}{
\Tr\big[ \exp \big( \frac{1}{4} \sum_{r,s} a_r W_{rs} a_s\big) \big]} \,,
\ee
with $W$ a complex antisymmetric matrix.
This density matrix is univocally identified by the correlation matrix 
\be
\label{Gamma def}
\Gamma_{rs} = \Tr (a_r \rho_W a_s) - \delta_{rs},
\ee
which   (in matrix form) satisfies $\Gamma =\tanh(W/2)$.

Unfortunately, the same reasoning does not apply to two disjoint blocks because  
the fermions in the interval $B_1$ separating the two blocks  
contribute to the spin reduced density matrix of $A_1\cup A_2$ \cite{atc-10,ip-10,fc-10}. 
In particular, the following string of Majorana operators appears in a crucial way\footnote{To compare with Ref. \cite{fc-10}, set 
$(S_1)_{\rm there} = (P_{A_1})_{\rm here}$ and $(S)_{\rm there} = (P_{B_1})_{\rm here}$.}
\be
\label{P_B1 def}
P_{B_1} \equiv \prod_{j\in B_1} ({\rm i} a_{2j-1} a_{2j}).
\ee
Similarly, we also introduce the strings of Majorana operators $P_{A_1}$ and $P_{A_2} $, defined as in Eq. (\ref{P_B1 def}) 
but taking the product over $j\in A_1$ and $j\in A_2$ respectively.
These operators satisfies $P_{C} ^{-1} = P_{C} $, for any $C=A_1,A_2,B_1$.


Moving back to the computation of the spin reduced density matrix (\ref{rhoA correlators}) in terms of fermions, 
we first notice that, since the XY Hamiltonian commutes with $\prod_j \sigma^z_j$ (where $j$ runs over the whole chain),
only the expectation values of operators containing an even number of fermions are non vanishing. 
Thus, the numbers of fermions are either even or odd in both $A_1$ and $A_2$.
This leads us to decompose the spin reduced density matrix $\rho_A$ of $A = A_1 \cup A_2$ as \cite{fc-10}
\be
\label{rhoA dec even-odd}
\rho_A 
\,=\, 
\rho_{\rm even} +  P_{B_1} \, \rho_{\rm odd} ,
\ee
with
\be\fl 
\label{rho even-odd def}
\rho_{\rm even} \equiv
\frac{1}{2^{\ell_1+\ell_2}}
\sum_{\rm even}
\langle O_1 O_2 \rangle\, O_2^\dagger O_1^\dagger ,
\qquad
\rho_{\rm odd}  \equiv
\frac{1}{2^{\ell_1+\ell_2}}
\sum_{\rm odd}
\langle O_1 P_{B_1}  O_2 \rangle\, O_2^\dagger  O_1^\dagger ,
\ee
where we introduced the `shorts' $O_1$ and $O_2$ for arbitrary products of Majorana fermions in $A_1$ and $A_2$ 
respectively, and the notation
$\sum_{\rm even}$ ($\sum_{\rm odd}$) means that the sum is restricted to operators $O_1$ and $O_2$ 
containing an even (odd) number of fermions. 
It is also convenient to rewrite (\ref{rhoA dec even-odd}) as
\be
\label{rho pm def}
\rho_A 
\,=\,
\frac{{\bf 1}+P_{B_1} }{2} \, \rho_+ +  \frac{{\bf 1}- P_{B_1} }{2} \, \rho_-,
\qquad
\rho_\pm \equiv
\rho_{\rm even} \pm \rho_{\rm odd} ,
\ee
where ${\bf 1}$ is the identity matrix and $({\bf 1} \pm P_{B_1} )/2$ are orthogonal projectors.
Moreover, the matrices $\rho_\pm$ are unitary equivalent (indeed $P_{A_2}  \rho_\pm P_{A_2}  = \rho_\mp$) and 
commute with $P_{B_1} $ because they do not contain Majorana fermions in $B_1$.
As a consequence, we have that
\be
\label{rhoA^n splitted}
\rho_A^n
\,=\,
\frac{{\bf 1}+P_{B_1} }{2} \, \rho_+^n +  \frac{{\bf 1}- P_{B_1} }{2} \, \rho_-^n.
\ee
Taking the trace of (\ref{rhoA^n splitted}) and employing that $P_{A_2}  \rho_\pm P_{A_2}  = \rho_\mp$, one finds
\be
\label{traces n rhopm}
\Tr \rho_A^n =  \Tr \rho_\pm^n .
\ee
The matrices $\rho_\pm$ are fermionic but they are not Gaussian, i.e. they are not proportional to the exponential of a quadratic form,
as we will discuss below.

At this point we are ready to consider the ground state of the XY chain with density matrix 
$\rho_W = |\Psi \rangle \langle \Psi |$, whose correlation matrix $\Gamma$ is given by Eq. (\ref{Gamma def}).
The fermionic reduced density matrix of $A=A_1 \cup A_2$  is
\be
\label{rhoA ff}
\rho_A^{\,{\bf 1}}
\equiv
\frac{1}{2^{\ell_1+\ell_2}}
\sum_{{\rm even} \atop {\rm odd}}
\langle O_1 O_2 \rangle\, O_2^\dagger O_1^\dagger ,
\ee
where we recall that $\langle O_1 O_2 \rangle = 0$ when the numbers of fermionic operators in $O_1$ and $O_2$ have different parity. 
In order to take into account the effect of the string $P_{B_1}$ defined in Eq. (\ref{P_B1 def}), 
it is useful to introduce the auxiliary density matrix
\be
\label{fake rhoA}
\rho_A^{B_1} \equiv \frac{\Tr_B \big(P_{B_1}  |\Psi \rangle \langle \Psi | \big)}{\langle P_{B_1}  \rangle},
\ee
in which the normalisation $\Tr  \rho_A^{B_1} =1 $ holds. 
By using that
\be
P_{A_2} a_j  P_{A_2}  = \bigg\{\begin{array}{rcc}
-\, a_j & & j\in A_2 ,\\
 a_j & & j\notin A_2, 
\end{array}
\ee
one finds
\bea
P_{A_2}  \rho_A^{\,{\bf 1}} P_{A_2} 
&=&
\frac{1}{2^{\ell_1+\ell_2}}
\sum_{{ {\rm even} \atop  {\rm odd}}}
(-1)^{\mu_2}
\langle O_1 O_2 \rangle\, O_2^\dagger O_1^\dagger 
\\
\label{PrhoP dec v2}
& =&
\frac{1}{2^{\ell_1+\ell_2}}
\sum_{\rm even}
\langle O_1 O_2 \rangle\, O_2^\dagger O_1^\dagger 
-
\frac{1}{2^{\ell_1+\ell_2}}
\sum_{\rm odd}
\langle O_1 O_2 \rangle\, O_2^\dagger O_1^\dagger ,
\eea
where $\mu_2$ is the number of Majorana operators occurring in $O_2$.
Then, from Eqs. (\ref{rhoA ff}) and (\ref{PrhoP dec v2}) it is straightforward to get that the density matrices (\ref{rho even-odd def}) become
\be
\label{rho pm ferm}
\rho_{\rm even}   =
\frac{\rho_A^{\,{\bf 1}} + P_{A_2}  \rho_A^{\,{\bf 1}} P_{A_2} }{2},
\qquad
\rho_{\rm odd}  =
\langle P_{B_1} \rangle
\,\frac{\rho_A^{B_1} - P_{A_2}  \rho_A^{B_1} P_{A_2} }{2}.
\ee

Plugging (\ref{rho pm ferm}) into (\ref{rhoA dec even-odd}), one finds that the spin reduced density matrix is a linear combination of 
four fermionic Gaussian operators.
Since these operators do not commute, they cannot be diagonalised simultaneously and therefore we cannot find the eigenvalues 
of the spin reduced density matrix that would give the entanglement entropy. 
Nevertheless, $\Tr \rho_A^n $ for integer $n$ can be computed through Eq. (\ref{traces n rhopm}) by providing the product rules between the 
four Gaussian operators occurring in (\ref{rho pm ferm}) in terms of the corresponding correlation matrices, 
that are denoted by 
\be
\Gamma_1 \equiv \Gamma_{\rho_A^{\,{\bf 1}} }, \qquad 
\Gamma_2 \equiv \Gamma_{P_{A_2}  \rho_A^{\,{\bf 1}} P_{A_2} }, \qquad
\Gamma_3 \equiv \Gamma_{\rho_A^{B_1}}, \qquad
\Gamma_4 \equiv \Gamma_{P_{A_2}  \rho_A^{B_1} P_{A_2} },
\ee
where $\Gamma_{\rho}$ is the correlation matrix of a Gaussian density matrix $\rho$ as in Eq. (\ref{Gamma def}).
Obviously, $\Gamma_1$ is the fermionic correlation matrix, i.e. the one of the free fermions 
without the Jordan-Wigner string (studied in detail in Ref. \cite{pascazio-08}).

Following Ref. \cite{fc-10}, we can introduce the restricted correlation matrix to two fermionic sets/blocks $C$ and $D$
$(\Gamma_{CD})_{rs} $ which is the correlation matrix in Eq. (\ref{Gamma def}) associated to 
$|\Psi \rangle \langle \Psi |$ with the restriction $r\in C $ and $s\in D$.
In \cite{fc-10} it has been shown that the matrices $\Gamma_2$, $\Gamma_3$ and $\Gamma_4$ can be written as
\be
\Gamma_1 = \Gamma_{AA},\qquad
\Gamma_3 =
\Gamma_1 - \Gamma_{AB_1}  \Gamma^{-1}_{B_1B_1}  \Gamma_{B_1 A} ,
\ee
and
\be
\label{Gamma24}
\Gamma_2  =  M_2 \Gamma_1 M_2,
\qquad
\Gamma_4 =  M_2 \Gamma_3 M_2,
\qquad
M_2 \equiv
\left(\begin{array}{cc}
{\bf 1}_{\ell_1} & {\bf 0}  \\   {\bf 0}  & - {\bf 1}_{\ell_2}
\end{array}\right).
\ee
These formulas provide an explicit representation of the matrices $\Gamma_i$ in terms of the 
fermion correlation in the finite subsystem $A_1\cup B_1\cup A_2$.

By introducing the following notation
\be
\label{Gamma prods notation}
\{ \dots , \Gamma, \dots , \Gamma', \dots \} 
\equiv 
\Tr (\dots \rho_W\dots  \rho_{W'} \dots ),
\ee
and $\{ \dots , \Gamma^n, \dots  \} \equiv \Tr (\dots \rho_W^n \dots )$ as special case, 
from (\ref{rho pm ferm}) we have that $ \Tr \rho_\pm^n $ can be written as a linear combination of  traces involving the matrices $\Gamma_k$ with $k\in \{1,2,3,4\}$. 
This finally provides ${\rm Tr} \rho_A^n $, which we write as follows
\be
\label{tr rho^n chain}
\textrm{Tr} \rho_A^n \equiv \frac{T_n}{2^{n-1}}.
\ee
From (\ref{traces n rhopm}) and (\ref{rho pm ferm}), it is straightforward to realise that $T_n$ is a combination of $4^n$ terms of the 
form (\ref{Gamma prods notation}) with coefficients given by integer powers of 
$\delta_{B_1} \equiv \langle P_{B_1} \rangle^2=\det [\Gamma_{B_1 B_1}]$.
However, many of these $4^n$ terms turn out to be equal when using cyclicality of the trace and other simple algebraic manipulations. 

In the following we write $T_n$ explicitly for $2\leqslant n \leqslant 5$, where we have computed also $T_5$, in addition to the other ones already reported in Ref. \cite{fc-10}:
\\
$\bullet$  $n=2$:
\be
T_2 \,=\,
\{ \Gamma_1^2 \}+\{ \Gamma_1 , \Gamma_2 \}
+ \delta_{B_1} \Big(
\{\Gamma_3^2  \} - \{ \Gamma_3 , \Gamma_4 \}
\Big);
\ee
$\bullet$  $n=3$:
\be
T_3 \,=\,
\{ \Gamma_1^3 \}+3\,\{ \Gamma_1^2 , \Gamma_2 \}
+ 3 \delta_{B_1}
 \Big(
\{\Gamma_1 , \Gamma_3^2  \}
+ \{\Gamma_2 , \Gamma_3^2  \}
- 2\,\{ \Gamma_1,\Gamma_4 , \Gamma_3 \}
\Big);
\ee
$\bullet$  $n=4$:
\bea
& &
\hspace{-2.5cm}
T_4 \,=\,
\{ \Gamma_1^4 \}
+ \{ \Gamma_1 , \Gamma_2 , \Gamma_1 , \Gamma_2 \}
+4\,\{ \Gamma_1^3 , \Gamma_2 \}
+2\,\{ \Gamma_1^2 , \Gamma_2^2 \}
\\
\rule{0pt}{.6cm}
& & \hspace{-1.5cm}
+2\delta_{B_1} \Big(
 \{ \Gamma_1 , \Gamma_3 , \Gamma_1 , \Gamma_3 \}
+\{ \Gamma_1 , \Gamma_4 , \Gamma_1 , \Gamma_4 \}
+2\, \{ \Gamma_1^2 , \Gamma_3^2  \}
+2\, \{ \Gamma_1^2 , \Gamma_4^2  \}
\nonumber\\
& & \hspace{.1cm}
+ 2\, \{ \Gamma_1 , \Gamma_3 , \Gamma_2 , \Gamma_3 \}
+4\, \{ \Gamma_1 , \Gamma_2 , \Gamma_3^2  \}
-2\big[ \,2\,\{  \Gamma_1^2, \Gamma_3 , \Gamma_4   \}
\nonumber\\
& &\hspace{.1cm}
+\{ \Gamma_1 , \Gamma_3 , \Gamma_1 , \Gamma_4 \}
+\{ \Gamma_1 , \Gamma_2 , \Gamma_3 , \Gamma_4  \}
+\{ \Gamma_1 , \Gamma_3 , \Gamma_2 , \Gamma_4  \}
+\{ \Gamma_1 , \Gamma_2 , \Gamma_4 , \Gamma_3  \}
\big]\Big)
\nonumber\\
\rule{0pt}{.1cm}
& &\hspace{-1.5cm}
+\delta_{B_1}^2  \Big(
 \{ \Gamma_3^4  \}
+2\,\{ \Gamma_3^2 , \Gamma_4^2 \}
+\{ \Gamma_3 , \Gamma_4,  \Gamma_3 , \Gamma_4 \}
-4\,\{ \Gamma_3^3 , \Gamma_4 \}
\Big);
\nonumber
\eea
$\bullet$  $n=5$
\bea
& &
\hspace{-2.5cm}
T_5 \,=\,
\TrProdNorm{5} + 
5 \,\Big( 
\TrProd{1}[4][2]
+\TrProd{1}[3][2][2]
+\TrProd{1}[2][2][][1][][2]
\Big)
\\
\rule{0pt}{.5cm}
& &
\hspace{-2.5cm}
+ 5\delta_{B_1} \Big( 
\TrProd{1}[3][3][2]
+\TrProd{1}[3][4][2]
+ 2 \,\TrProd{1}[][2][2][3][2]
+ 2 \,\TrProd{1}[2][2][][3][2]
\nonumber \\
& &
\hspace{-2cm}
+ \TrProd{1}[][2][][1][][3][2]
+ \TrProd{1}[][2][][1][][4][2]
+ \TrProd{1}[2][3][][1][][3]
+ \TrProd{1}[2][3][][2][][3]
\nonumber \\
& &
\hspace{-2cm}
+ 2 \,\TrProd{1}[][3][][1][][3][][2]
+ \TrProd{1}[][3][][2][2][3]
+ \TrProd{1}[2][4][][1][][4]
+ 2\,\TrProd{1}[][2][][3][][2][][3]
\nonumber \\
& &
\hspace{-2cm}
- 2\big[ \TrProd{1}[2][2][][4][][3]
+ \TrProd{1}[2][2][][3][][4]
+\TrProd{1}[2][3][][1][][4]
+\TrProd{1}[2][3][][2][][4]
\nonumber \\
& &
\hspace{-1.1cm}
+\TrProd{1}[3][3][][4]
+\TrProd{1}[][2][][1][][3][][4]
+ \TrProd{1}[][2][][4][][1][][3]
+\TrProd{2}[][1][][4][][1][][3] \big]
\Big)
\nonumber \\
& &
\rule{0pt}{.5cm}
\hspace{-2.5cm}
+ 5\delta_{B_1}^2 \Big( 
\TrProd{1}[][3][4]
+ \TrProd{2}[][3][4]
+ 2 \,\TrProd{1}[][3][2][4][2]
+ \TrProd{1}[][3][][4][2][3]
+ 2 \,\TrProd{1}[][3][][4][][3][][4]
\nonumber \\
& &
\hspace{-2.1cm}
+ \TrProd{1}[][4][][3][2][4]
- 2\big[
 \TrProd{1}[][3][3][4]
+ \TrProd{1}[][3][2][4][][3]
+ \TrProd{2}[][4][][3][3]
+ \TrProd{2}[][3][2][4][][3]
\big] \Big).
\nonumber 
\eea

We notice that the algebraic sum of the integer coefficients occurring in any term multiplying a power $\delta_{B_1}^p$ with $p>0$ is zero. 
Moreover, considering only the terms which are not multiplied by a power $\delta_{B_1}$ in $T_n$, the sum of their coefficients is $2^{n-1}$.


\section{Traces of integer powers of the partial transpose of the spin reduced density matrix}
\label{sec pt2}

In this section we move to the main objective of this paper which is to give a representation of the integer powers of the partial 
transpose of the spin reduced density matrix of two disjoint blocks with respect to $A_2$.
Eisler and Zimboras in Ref. \cite{ez-15} showed how to obtain the partial transpose of a fermionic Gaussian 
density matrix, a procedure which can be applied to the spin reduced density matrix in Eq. (\ref{rho pm def}) using the linearity 
of the partial transpose as we are going to show. 
We mention that in Ref. \cite{ez-15}  the moments of the partial transpose for two adjacent intervals were studied in 
details using the property that fermionic and spin reduced density matrices are equal for this special case. 

Given a Gaussian density matrix $\rho_W$ written in terms of Majorana fermions in $A=A_1 \cup A_2$, the partial transposition 
with respect to $A_2$ leaves invariant the modes in $A_1$ and acts only on the ones in $A_2$.
Furthermore, the partial transposition with respect to $A_2$ of $\rho_A$ in (\ref{rhoA dec even-odd}) leaves the operator $P_{B_1}$  
unchanged (because it does not contain modes in $A_2$), therefore we have
\be
\label{rhoA T2}
\rho_A^{T_2} 
\,=\, 
\rho_{\rm even}^{T_2} +  P_{B_1} \, \rho_{\rm odd}^{T_2} 
\,=\,
\frac{{\bf 1}+P_{B_1} }{2} \, \rho_+^{T_2} 
+  \frac{{\bf 1}- P_{B_1} }{2} \, \rho_-^{T_2},
\ee
where 
\be
\label{rho pm T2}
\rho_\pm^{T_2} =
\rho_{\rm even}^{T_2} \pm \rho_{\rm odd}^{T_2}, 
\ee
as clear from Eq. (\ref{rho pm def}) because of the linearity of the partial transpose.
The partial transposition of an arbitrary product of Majorana fermions $A_2$ (denoted shortly as $O_2$ like in the previous section)
is given by the following map \cite{ez-15}
\be
\label{R2 ez def}
\mathcal{R}_2(O_2) \equiv (-1)^{\tau(\mu_2)} O_2,
\qquad
\tau(\mu_2) \equiv \bigg\{\begin{array}{ll}
0 \;\;& (\mu_2 \textrm{ mod } 4) \in \{0,3\},
\\
1 & (\mu_2 \textrm{ mod } 4) \in \{1,2\},
\end{array}
\ee
where we recall that $\mu_2$ is the number of Majorana operators in $O_2$.
Then,  applying Eq. (\ref{R2 ez def}) to (\ref{rho even-odd def}), we find
\bea
\rho_{\rm even}^{T_2} =
\frac{1}{2^{\ell_1+\ell_2}}
\sum_{\rm even}
(-1)^{\mu_2/2}
\langle O_1 O_2 \rangle\, O_2^\dagger O_1^\dagger ,
\nonumber\\
\rho_{\rm odd}^{T_2} =
\frac{1}{2^{\ell_1+\ell_2}}
\sum_{\rm odd}
(-1)^{(\mu_2-1)/2}
\langle O_1 P_{B_1}  O_2 \rangle\, O_2^\dagger  O_1^\dagger ,
\label{rho T2 even-odd}
\eea
which gives the desired fermionic representation of the partial transpose of the spin reduced density matrix. 

At this point the moments of $\rho^{T_2}_A$ can be obtained following the same reasoning as for the moments of $\rho_A$.
Indeed, since $\rho_\pm^{T_2}$ are unitarily equivalent ($P_{A_2}  \rho_\pm^{T_2} P_{A_2}  = \rho_\mp^{T_2} $ 
because $P_{A_2} \rho_{\rm even}^{T_2} P_{A_2} = \rho_{\rm even}^{T_2}$ and 
$P_{A_2} \rho_{\rm odd}^{T_2} P_{A_2} = - \rho_{\rm odd}^{T_2}$) and $P_{B_1}$ 
commutes with them, starting from (\ref{rhoA T2}) 
and repeating the same observations that lead to (\ref{traces n rhopm}), one gets
\be
\label{traces n rhopm T2}
\Tr \big( \rho_A^{T_2}\big)^n 
=  \Tr \big(\rho_\pm^{T_2}\big)^n .
\ee

Similarly to the case of the R\'enyi entropies  considered in Sec. \ref{sec renyi} 
(see Eq. (\ref{traces n rhopm})), the matrices $\rho_\pm^{T_2}$ are fermionic but not Gaussian.
In the following we write them as sums of four Gaussian matrices, as done in (\ref{rhoA dec even-odd}) and (\ref{rho pm ferm}) for $\rho_A$.
In particular, by introducing
\be
\fl 
\label{rhoA tilde B1 def}
\tilde{\rho}_A^{\,{\bf 1}} \equiv
\frac{1}{2^{\ell_1+\ell_2}}
\sum_{{\rm even} \atop {\rm odd}}
\textrm{i}^{\mu_2}
\langle O_1 O_2 \rangle\, O_2^\dagger O_1^\dagger ,
\qquad
\tilde{\rho}_A^{B_1} \equiv
\frac{1}{2^{\ell_1+\ell_2}}
\sum_{{\rm even} \atop {\rm odd}}
\textrm{i}^{\mu_2}\,
\frac{\langle O_1 P_{B_1}  O_2 \rangle}{\langle P_{B_1}  \rangle}
\; O_2^\dagger O_1^\dagger ,
\ee
one has that the matrices in (\ref{rho T2 even-odd}) become
\be
\label{rhoT2 even-odd}
\rho_{\rm even}^{T_2} =
\frac{\tilde{\rho}_A^{\,{\bf 1}} + P_{A_2}  \tilde{\rho}_A^{\,{\bf 1}} P_{A_2} }{2},
\qquad
\rho_{\rm odd}^{T_2}  =
\langle P_{B_1} \rangle\,
\frac{\tilde{\rho}_A^{B_1} - P_{A_2}  \tilde{\rho}_A^{B_1} P_{A_2} }{2 \textrm{i}},
\ee
telling us that $\rho_\pm^{T_2}$ in (\ref{rho pm T2}) are linear combinations of four Gaussian fermionic matrices 
occurring in the r.h.s.'s of (\ref{rhoT2 even-odd}).
Notice that $\rho_{\rm even}^{T_2}$ and $\rho_{\rm odd}^{T_2} $ are Hermitian but the matrices defining them are not since
\be
\big(\tilde{\rho}_A^{\,{\bf 1}}\big)^\dagger 
= P_{A_2}  \tilde{\rho}_A^{\,{\bf 1}} P_{A_2} ,
\qquad
\big(\tilde{\rho}_A^{B_1}\big)^\dagger =
P_{A_2}  \tilde{\rho}_A^{B_1} P_{A_2} .
\ee

In order to compute the correlation matrices associated to the four matrices in Eq. (\ref{rhoT2 even-odd}), it is convenient to introduce
\be
\widetilde{M}_2 \equiv
\left(\begin{array}{cc}
{\bf 1}_{\ell_1} & {\bf 0}  \\  {\bf 0}  & \textrm{i} {\bf 1}_{\ell_2}
\end{array}\right).
\ee
Then, the correlation matrices associated to $\tilde{\rho}_A^{\,{\bf 1}} $, $P_{A_2}  \tilde{\rho}_A^{\,{\bf 1}} P_{A_2}$, $\tilde{\rho}_A^{B_1} $ and $P_{A_2}  \tilde{\rho}_A^{B_1} P_{A_2} $ are given  by
\be
\widetilde\Gamma_k  \equiv  
\widetilde{M}_2  \Gamma_k \widetilde{M}_2,
\qquad
k\,\in\,\{1,2,3,4\}.
\ee
In analogy to Eq. (\ref{tr rho^n chain}), we write the moments of $\rho^{T_2}_A$ as 
\be
\label{tr rho^n chain T2}
\Tr \big(\rho^{T_2}_A\big)^n 
= 
\frac{\widetilde{T}_n}{2^{n-1}}.
\ee
From Eqs. (\ref{rho pm T2}), (\ref{traces n rhopm T2}) and (\ref{rhoT2 even-odd}), we have that $\widetilde{T}_n$ 
is a linear combination of $4^n$ terms. 
%
The net effect is that  $\widetilde{T}_n$ can be written by taking $T_n$ and replacing $\Gamma_i$ with 
$\widetilde\Gamma_i$ and $\delta_{B_1}$ with $- \delta_{B_1}$.
The latter rule comes from the imaginary unit in the denominator of $\rho_{\rm odd}^{T_2} $ in Eq. (\ref{rhoT2 even-odd}).

In the following we write explicitly $\widetilde{T}_n$ for $2 \leqslant n \leqslant 5$:\\
$\bullet$ $n=2$
\be
\label{T2tilde}
\widetilde{T}_2 \,=\,
\{ \widetilde{\Gamma}_1^2 \}
+ \{\widetilde{\Gamma}_1 ,\widetilde{\Gamma}_2 \}
+ \delta_{B_1} \Big( 
 \{ \widetilde{\Gamma}_3 , \widetilde{\Gamma}_4 \}
- \{\widetilde{\Gamma}_3^2  \}
\Big);
\ee
$\bullet$ $n=3$
\be
\label{T3tilde}
\widetilde{T}_3 \;=\;
\{ \widetilde{\Gamma}_1^3 \}
+3\,\{ \widetilde{\Gamma}_1^2 , \widetilde{\Gamma}_2 \}
+ 3\delta_{B_1}
 \Big(
 2\, \{ \widetilde{\Gamma}_1,\widetilde{\Gamma}_4 , \widetilde{\Gamma}_3 \}
- \{\widetilde{\Gamma}_1 , \widetilde{\Gamma}_3^2  \}
- \{\widetilde{\Gamma}_2 , \widetilde{\Gamma}_3^2  \}
\Big);
\ee
$\bullet$ $n=4$
\bea
\label{T4tilde}
& &
\hspace{-2.5cm}
\widetilde{T}_4 \,=\,
\{ \widetilde{\Gamma}_1^4 \}
+ \{ \widetilde{\Gamma}_1 , \widetilde{\Gamma}_2 , \widetilde{\Gamma}_1 , \widetilde{\Gamma}_2 \}
+4\,\{ \widetilde{\Gamma}_1^3 , \widetilde{\Gamma}_2 \}
+2\,\{ \widetilde{\Gamma}_1^2 , \widetilde{\Gamma}_2^2 \}
\\
\rule{0pt}{.5cm}
& &\hspace{-2.5cm}
+\,2\delta_{B_1} \Big(
2\, \{ \widetilde{\Gamma}_3 , \widetilde{\Gamma}_1 , \widetilde{\Gamma}_4 , \widetilde{\Gamma}_1 \}
+ 2\, \{ \widetilde{\Gamma}_1 , \widetilde{\Gamma}_2 , \widetilde{\Gamma}_3 , \widetilde{\Gamma}_4 \}
+ 2\, \{ \widetilde{\Gamma}_1 , \widetilde{\Gamma}_3 , \widetilde{\Gamma}_2 , \widetilde{\Gamma}_4 \}
+ 2\, \{ \widetilde{\Gamma}_1 , \widetilde{\Gamma}_2 , \widetilde{\Gamma}_4 , \widetilde{\Gamma}_3 \}
\nonumber
\\
& &\hspace{-.5cm}
+ 4\,\{ \widetilde{\Gamma}_3 , \widetilde{\Gamma}_4 , \widetilde{\Gamma}_1^2  \}
- 2\, \{ \widetilde{\Gamma}_1 , \widetilde{\Gamma}_3 , \widetilde{\Gamma}_2 , \widetilde{\Gamma}_3 \}
-4\,\{ \widetilde{\Gamma}_1 , \widetilde{\Gamma}_2 , \widetilde{\Gamma}_3^2  \} 
\nonumber \\
& &\hspace{-.5cm}
-  \{ \widetilde{\Gamma}_1 , \widetilde{\Gamma}_3 , \widetilde{\Gamma}_1 , \widetilde{\Gamma}_3 \}
- \{ \widetilde{\Gamma}_1 , \widetilde{\Gamma}_4 , \widetilde{\Gamma}_1 , \widetilde{\Gamma}_4 \}
-2\, \{ \widetilde{\Gamma}_1^2 , \widetilde{\Gamma}_3^2  \}
-2\, \{ \widetilde{\Gamma}_1^2 , \widetilde{\Gamma}_4^2  \}
\Big)
\nonumber\\
\rule{0pt}{.5cm}
& &\hspace{-2.5cm}
+\,\delta_{B_1}^2  \left(
\{ \widetilde{\Gamma}_3^4  \}
+2\,\{ \widetilde{\Gamma}_3^2 , \widetilde{\Gamma}_4^2 \}
+ \{ \widetilde{\Gamma}_3 , \widetilde{\Gamma}_4,  \widetilde{\Gamma}_3 , \widetilde{\Gamma}_4 \}
-4\,\{ \widetilde{\Gamma}_3^3 , \widetilde{\Gamma}_4 \}
\right);
\nonumber 
\end{eqnarray}
$\bullet$ $n=5$
\label{T5tilde}
\bea
& & \hspace{-2.5cm}
\widetilde{T}_5 \,=\,
\TrProdNormTilde{5} + 5 \Big( 
\TrProdTilde{1}[4][2]
+\TrProdTilde{1}[3][2][2]
+\TrProdTilde{1}[2][2][][1][][2]
\Big)
\\
\rule{0pt}{.5cm}
& &
\hspace{-2.5cm}
+ 5\delta_{B_1} \Big( 
2\,\TrProdTilde{1}[2][2][][4][][3]
+ 2\,\TrProdTilde{1}[2][2][][3][][4]
+ 2\,\TrProdTilde{1}[2][3][][1][][4]
+ 2\,\TrProdTilde{1}[2][3][][2][][4]
\nonumber \\
& & \hspace{-1.8cm}
+ 2\,\TrProdTilde{1}[3][3][][4]
+ 2\,\TrProdTilde{1}[][2][][1][][3][][4]
+ 2\,\TrProdTilde{1}[][2][][4][][1][][3]
+ 2\,\TrProdTilde{1}[][4][][1][][3][][2]
\nonumber \\
& & \hspace{-1.8cm}
-\,\TrProdTilde{1}[3][3][2]
- \TrProdTilde{1}[3][4][2]
- 2 \,\TrProdTilde{1}[][2][2][3][2]
- 2 \,\TrProdTilde{1}[2][2][][3][2]
\nonumber \\
& & \hspace{-1.8cm}
- \TrProdTilde{1}[][2][][1][][3][2]
- \TrProdTilde{1}[][2][][1][][4][2]
- \TrProdTilde{1}[2][3][][1][][3]
- \TrProdTilde{1}[2][3][][2][][3]
\nonumber \\
& & \hspace{-1.8cm}
-\, 2 \,\TrProdTilde{1}[][3][][1][][3][][2]
- \TrProdTilde{1}[][3][][2][2][3]
- \TrProdTilde{1}[2][4][][1][][4]
- 2 \,\TrProdTilde{1}[][2][][3][][2][][3]
\,\Big)
\nonumber \\
\rule{0pt}{.5cm}
& & \hspace{-2.5cm}
+ 5\delta_{B_1}^2 \left( 
\TrProdTilde{1}[][3][4]
+ \TrProdTilde{2}[][3][4]
+ \TrProdTilde{1}[][3][][4][2][3]
+ \TrProdTilde{1}[][4][][3][2][4]
+ 2 \,\TrProdTilde{1}[][3][][4][][3][][4]
\right. 
\nonumber \\
& & \hspace{-2.5cm}
+ 2\, \TrProdTilde{1}[][3][2][4][2]
- 2\,
 \TrProdTilde{1}[][3][3][4]
- 2\, \TrProdTilde{1}[][3][2][4][][3]
- 2\, \TrProdTilde{2}[][4][][3][3]
- 2\, \TrProdTilde{2}[][3][2][4][][3]
 \Big).
 \nonumber 
\eea
As for $T_n$, also in $\widetilde T_n$  the algebraic sum of the integer coefficients occurring in any term multiplying a power 
$\delta_{B_1}^p$ with $p>0$ vanishes.

\section{Numerical results for the ground state of the critical Ising and XX model}
\label{sec num}

The results of the previous section for the moments of the partial transpose of the reduced density matrix of two disjoint 
blocks are valid for arbitrary configurations of the XY spin chain: equilibrium, non-equilibrium, finite and infinite systems, 
critical and non-critical values of the parameters $\gamma$ and $h$.
In this section we evaluate numerically these moments for the configurations that so far attracted most of the theoretical interest,
namely the critical points of the XY Hamiltonian, whose scaling properties are described by conformal field theories.
A great advantage of the present approach compared to purely numerical methods such as 
exact diagonalization or tensor networks techniques is that it allows to deal directly with infinite chains without
any approximations, reducing the systematic errors in the estimates of asymptotic results. 
Indeed, all the numerical results presented in the following are obtained for infinite chains. 

We will consider two particular points of the XY Hamiltonian, 
namely the critical Ising model for $\gamma=h=1$ and the zero field XX spin chain 
(corresponding to fermions at half-filling) obtained for $\gamma=h=0$.
The scaling limit of the former is the Ising CFT with central charge $c=1/2$,
while the scaling limit of the latter is a compactified boson at the Dirac point with $c=1$.

The CFT predictions for the moments of both reduced density matrix and its partial transpose 
have been derived in a series of manuscripts and they are reviewed  in Appendix \ref{app1}. 
For both models we consider the case of two disjoint blocks of equal length $\ell$ embedded in an infinite 
chain and placed at distance $r$.
We numerically evaluate the moments of $\rho_A$ and $\rho_A^{T_2}$ using the trace 
formulas of the previous sections for $n=2,3,4,5$ and we compute 
the ratio $R_n$ (defined in Eq. (\ref{R_n def})):
\be
\label{R_n def2}
R_n 
\equiv \frac{\Tr (\rho^{T_2}_A)^n }{\Tr \rho_A^n },
\ee
whose (unknown) analytic continuation for $n_e\to1$ would give the negativity. 
Notice that from Eqs. (\ref{tr rho^n chain}) and (\ref{tr rho^n chain T2}) 
we have that $R_n = \widetilde{T}_n/T_n$.
In the scaling limit (i.e. $\ell,r\to\infty$ with ratio fixed) the ratio $R_n$ 
converges to the CFT prediction (cf. Eq. (\ref{RnCFT}) in Appendix) written in terms of the 
four-point ratio $x$, which is 
\be
x=\left(\frac{\ell}{\ell+r}\right)^2\,,
\label{4pRlat}
\ee
when specialised to the case of two intervals of equal length $\ell$ at distance $r$.

\subsection{The critical Ising chain}

\begin{figure}
\includegraphics[width=0.95\textwidth]{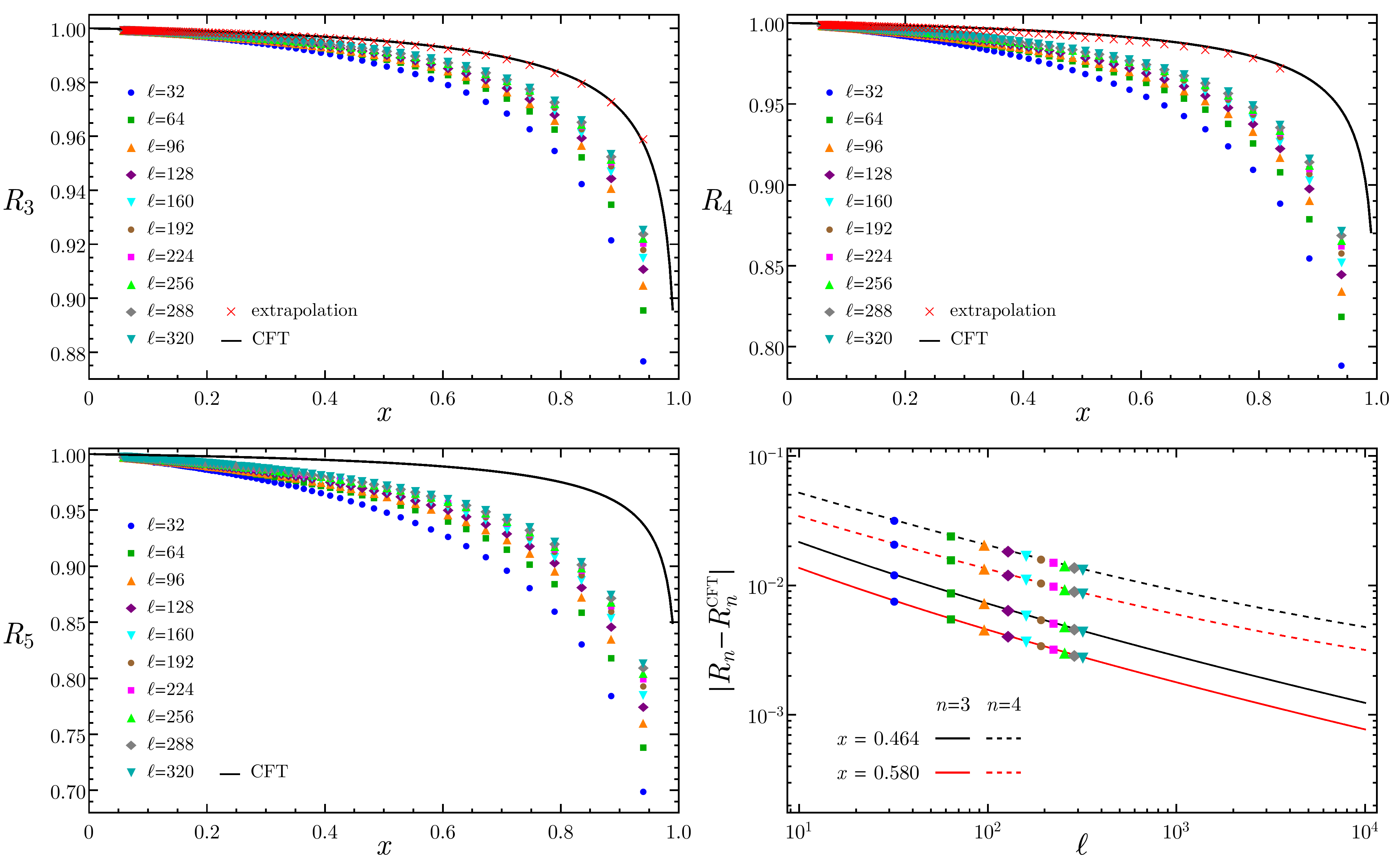}
\caption{The ratio $R_n$ between the integer moments of $\rho_A$ and $\rho_A^{T_2}$ for two 
disjoint blocks of length $\ell$ at distance $r$ embedded in an infinite critical Ising chain. 
We report the results for $n=3,4,5$ as function of the four-point ratio $x$ for various values of $\ell$ (and correspondingly of $r$).
For large $\ell$, the data approach the CFT predictions (solid lines).
The extrapolations to $\ell\to\infty$ --done using the scaling form (\ref{ansatz2R})-- are shown as crosses and they perfectly agree
with the CFT curves for $n=3$ and $4$, while for $n=5$ the fits are unstable and the extrapolations are not shown.
The last panel shows explicitly the extrapolating functions for two values of $x$ and $n=3,4$.
} 
\label{fig:IS}
\end{figure}

The negativity and the moments of $\rho^{T_2}_A$ for the critical Ising chain in a transverse field 
have been already numerically considered in Ref. \cite{ctt-13} by using
a tree tensor network algorithm and in Ref. \cite{a-13} by Monte Carlo simulations of the two-dimensional 
classical problem in the same universality class. 
However, the finiteness of the chain length did not allow to obtain very precise extrapolations 
to the scaling theory for all values of $n$ and of the four-point ratio $x$. 
We found, as generally proved \cite{cct-neg-letter}, that $R_2$ is identically equal to $1$.
In Fig. \ref{fig:IS} we report the obtained values of $R_n$ for $n=3,4,5$ as function of $x$ 
for different values of $\ell$.
It is evident that increasing $\ell$ the data approach the CFT predictions (the solid curves).
We can also perform an accurate scaling analysis to show that indeed the data converge to
the CFT results when the corrections to the scaling are properly taken into account. 

It has been argued  on the basis of the general CFT arguments \cite{cc-10}, 
and shown explicitly in few examples \cite{ce-10,ccen-10,un-vari} both analytically and numerically, that $\Tr \rho_A^n$ 
displays `unusual' corrections to the scaling which, at the leading order, 
are governed by  the unusual exponent $\delta_n=2h/n$ where $h$ is the smallest scaling dimension of 
a relevant operator which is inserted locally at the branch point \cite{cc-10}.
For the Ising model it has been found that, in the case of two intervals, $h=1/2$ \cite{atc-10,fc-10}. 
From the general CFT arguments in Ref. \cite{cc-10}, we expect the same corrections to be 
present for $\Tr (\rho_A^{T_2})^n$ because they are only due to the conical singularities.
Unfortunately, the corrections to the scaling in Fig. \ref{fig:IS} cannot be captured by a single term, 
because subleading corrections become more and more important when $n$ increases, as already pointed out  in Ref. \cite{ctt-13}. 
Indeed, corrections of the form  $\ell^{-m/n}$  for any integer $m$ are know to be present \cite{ce-10,fc-10,atc-11}. 
Thus  the most general finite-$\ell$ ansatz is of the form
\be
R_n=R_n^{\rm CFT}(x)+\frac{r_n^{(1)}(x)}{\ell^{1/n}} +\frac{r_n^{(2)}(x)}{\ell^{2/n}}+\frac{r_n^{(3)}(x)}{\ell^{3/n}}+\cdots\,.
\label{ansatz2R}
\ee
The variables $r_n(x)$ are used as fitting parameters in the extrapolation procedure. 
The number of terms that we should keep in order to have a stable fit depends both on $n$ and on $x$.
For each case we keep a number of terms such that the extrapolated value at $\ell\to\infty$ 
is stable. In any case we never keep corrections beyond the order $O(\ell^{-1})$.
The results of this extrapolation procedure for $n=3$ and $4$ are explicitly reported in Fig. \ref{fig:IS}.
The agreement of the extrapolations with the CFT predictions is really excellent, at an unprecedented precision 
compared with fully numerical computations \cite{ctt-13,a-13}.
Conversely, we find that for $n=5$ the extrapolations are still unstable because of the large number of terms 
we should keep in order to have a precise enough extrapolation.

\subsection{The XX chain}

\begin{figure}
\includegraphics[width=0.95\textwidth]{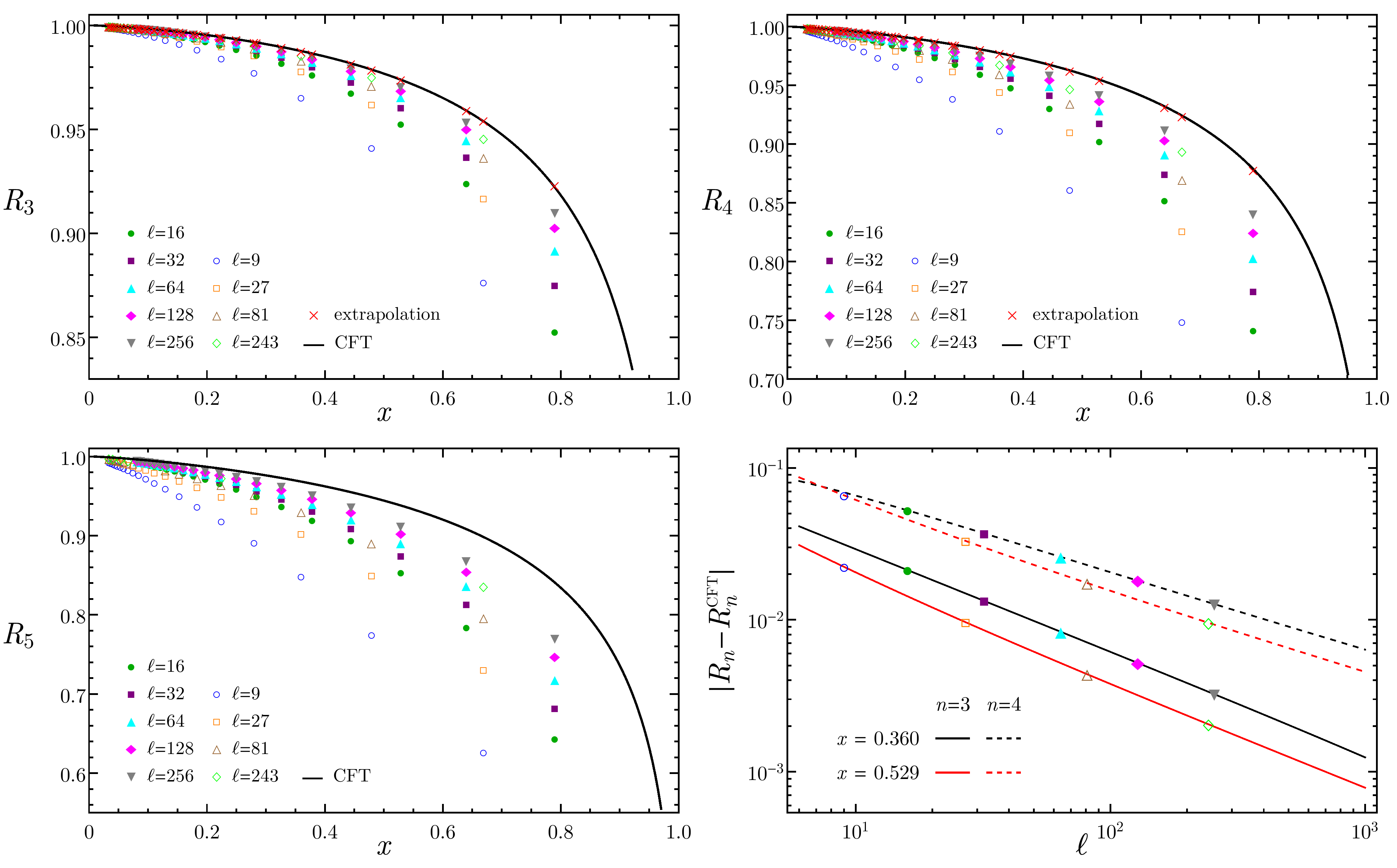}
\caption{The ratio $R_n$ between the integer moments of $\rho_A$ and $\rho_A^{T_2}$ for two 
disjoint blocks of length $\ell$ at distance $r$ embedded in an infinite XX chain at zero field. 
We report the results for $n=3,4,5$ as function of the four-point ratio $x$ for various values of $\ell$ (and correspondingly of $r$).
For large $\ell$, the data approach the CFT predictions (solid lines).
The extrapolations to $\ell\to\infty$ --done using the scaling form (\ref{ansatz2R2})-- are shown as crosses which perfectly agree
with the CFT curves for $n=3$ and $4$, while for $n=5$ the fits are unstable.
The last panel shows explicitly the extrapolating functions for two values of $x$ and $n=3,4$.
} 
\label{fig:XX}
\end{figure}

We now move to the study of the powers of $\rho_A^{T_2}$ for the XX model in zero field. 
There are no previous numerical studies of this paradigmatic model. 
We again consider the ratios $R_n$ for $n=2,3,4,5$ and we again find  
that $R_2$ is identically equal to $1$, as it should be.
In Fig. \ref{fig:XX} we report the obtained values of $R_n$ for $n=3,4,5$ as function of $x$ 
for different values of $\ell$.
It is evident that increasing $\ell$ the data approach the CFT predictions (the solid curves).
We should however mention a very remarkable property. 
It has been observed that $\Tr\rho_A^n$ shows oscillating corrections to the scaling \cite{ce-10,ccen-10,fc-10},
which for zero magnetic field, are of the form $(-1)^\ell$.
These oscillations however cancel in the ratio $R_n$ and the corrections to the scaling are 
monotonous, a property which makes the extrapolation to infinite $\ell$ slightly simpler.

Also in this case we can perform an accurate scaling analysis to show how the data converge to
the CFT results when the corrections to the scaling are properly taken into account. 
For the XX model, the leading correction to the scaling is governed by an exponent $\delta_n=2/n$, 
which means that they are less severe than in the case of the Ising model as it is also qualitatively clear from the figure. 
We then use the general finite-$\ell$ ansatz 
\be
R_n=R_n^{\rm CFT}(x)+\frac{r_n^{(1)}(x)}{\ell^{2/n}} +\frac{r_n^{(2)}(x)}{\ell^{4/n}}+\frac{r_n^{(3)}(x)}{\ell^{6/n}}+\cdots\,,
\label{ansatz2R2}
\ee
and, as in the case of the Ising model, we keep a number of fitting parameters which make stable the extrapolation 
at $\ell\to\infty$.
The results of this procedure for $n=3$ and $4$ are explicitly reported in Fig. \ref{fig:XX}.
The agreement of the extrapolations with the CFT predictions is excellent.
Also for the XX chain we find that for $n=5$ the fits are unstable.

\section{Two disjoint intervals for free fermions}
\label{sec FF}

In this section we consider the partial transposition for two disjoint blocks in the fermionic variables.
This problem was already addressed by Eisler and Zimboras \cite{ez-15}, but a detailed numerical analysis 
was not presented.
For fermionic variables there is no string in $B_1$ connecting the two blocks (cf. Eq. (\ref{P_B1 def})).
Thus the partial transpose of fermions can be obtained from the formulas derived 
in the previous sections by discarding the string of Majorana operators (\ref{P_B1 def}), i.e. by replacing 
$P_{B_1}$ with ${\bf 1}$. Performing this replacement, many simplifications occur in the formulas found in Sec. \ref{sec renyi} and 
Sec. \ref{sec pt2} as we will discuss in the following.

\subsection{R\'enyi entropies}

By definition the  fermionic reduced density matrix is Gaussian with correlation matrix $\Gamma_1$ defined in the 
Sec. \ref{sec renyi}, i.e.
\be
\Tr \rho_A^n=\{\Gamma_1^n\}.
\label{momff}
\ee
It is however instructive to recover this result from the formulas in Sec. \ref{sec renyi} in order to set up the calculation 
for the partial transpose. 

Making the replacement $P_{B_1} \to {\bf 1}$, the reduced density matrix of the two  disjoint blocks given in 
Eqs. (\ref{rhoA dec even-odd}) and (\ref{rho even-odd def}) becomes
\be
\label{rhoA dec even-odd free}
\rho_A 
\,=\, 
\rho_{\rm even} +  \rho_{\rm odd}^F,
\ee
where\footnote{To compare our notation with the one used in \cite{ez-15}, set $(\rho_+)_{\rm there} = (\rho_{\rm even})_{\rm here}$ 
and $(\rho_-)_{\rm there} = (\rho_{\rm odd}^F)_{\rm here}$.}
\be\fl 
\label{rho even-odd def free}
\rho_{\rm even} =
\frac{1}{2^{\ell_1+\ell_2}}
\sum_{\rm even}
\langle O_1 O_2 \rangle\, O_2^\dagger O_1^\dagger ,
\qquad
 \rho_{\rm odd}^F
=
\frac{1}{2^{\ell_1+\ell_2}}
\sum_{\rm  odd}
\langle O_1  O_2 \rangle\, O_2^\dagger  O_1^\dagger .
\ee
Moreover, $\rho_A^{B_1}$ defined in Eq. (\ref{fake rhoA}) is replaced as $\rho_A^{B_1} \to\rho_A$ and therefore, 
from Eqs. (\ref{rho pm ferm}) and (\ref{rhoA dec even-odd free}) we conclude that $\rho_+^F =  \rho_A^{\,{\bf 1}} $.
As for the correlation matrices $\Gamma_i$, since $\rho_A^{B_1} \to\rho_A$, it is obvious that 
$\Gamma_3 \to \Gamma_1$ and $\Gamma_4 \to \Gamma_2$.

Summarising, we conclude that the fermionic $\Tr \rho_A^n$ 
is found by making  in Eq. (\ref{tr rho^n chain}) the following replacements
\be
\delta_{B_1} \to 1,
\qquad
\Gamma_3 \to \Gamma_1,
\qquad
\Gamma_4 \to \Gamma_2.
\ee
Performing these substitutions in the explicit examples given in Sec. \ref{sec renyi} for $2 \leqslant n \leqslant 5$, 
it is straightforward to find Eq. (\ref{momff}),
which is just the obvious result that the fermionic density matrix is the Gaussian operator with 
correlation matrix given by $\Gamma_1$.

As a further check of our numerical codes, we numerically calculated $\Tr\rho_A^n$ using Eq. (\ref{momff})
(as was already done in Ref. \cite{pascazio-08}), obtaining that on the critical lines in the scaling limit 
it converges to  
\be
\Tr\rho_A^n\to\,
  \frac{c_n^2}{\big[\ell_1\ell_2(1-x)\big]^{2\Delta_n}},
\ee
that corresponds to ${\cal F}_n(x)=1$ identically in the general CFT formula (\ref{tr rhoAn cft}).
Indeed this result was already proven in the continuum free fermion theory \cite{ch-05}.

\subsection{Traces of integer powers of the partial transpose}

We are now ready to set up the formulas for the moments of the partial transpose, 
as already derived by Eisler and Zimboras \cite{ez-15}, but numerically studied only for 
the case of adjacent intervals.

Once again, $\Tr (\rho^{T_2}_A)^n$ in the fermionic variables is obtained by replacing  $P_{B_1}$ with ${\bf 1}$ 
in the formulas reported in Sec. \ref{sec pt2}.
Performing this replacement in Eq. (\ref{rhoA T2}) 
we get $\rho_A^{T_2} =\rho_{\rm even}^{T_2} +  (\rho_{\rm odd}^F)^{T_2}
\,=\, \rho_+^{T_2} $ and in Eq. (\ref{rhoA tilde B1 def}) it gives $\tilde{\rho}_A^{B_1}  \to \tilde{\rho}_A^{\,{\bf 1}} $. 
These observations together with Eq. (\ref{rhoT2 even-odd}) lead to
\be
\rho_A^{T_2}  = 
\frac{1-\textrm{i}}{2}\, \tilde{\rho}_A^{\,{\bf 1}} 
+
\frac{1+\textrm{i}}{2}\,  P_{A_2}  \tilde{\rho}_A^{\,{\bf 1}} P_{A_2},
\ee
which is exactly the same result obtained in \cite{ez-15} 
(for direct comparison, set $(O_+)_{\rm there} = (\tilde{\rho}_A^{\,{\bf 1}} )_{\rm here}$ 
and $(O_-)_{\rm there} = (P_{A_2}  \tilde{\rho}_A^{\,{\bf 1}} P_{A_2})_{\rm here}$).

The last remaining step is just to write these formulas in terms of correlation matrices.
Given that $\tilde{\rho}_A^{B_1}  \to \tilde{\rho}_A^{\,{\bf 1}} $, it follows that we should perform the replacements 
$\widetilde{\Gamma}_3 \to \widetilde{\Gamma}_1$ and $\widetilde{\Gamma}_4 \to \widetilde{\Gamma}_2$ 
in order to get the moments of the partial transpose in terms of the correlation matrices.  
Summarising, the fermionic $\Tr (\rho^{T_2}_A)^n$ are given by the formulas in Sec. \ref{sec pt2}
performing the replacements 
\be
\label{ff T2 rule}
\delta_{B_1} \to 1,
\qquad
\widetilde{\Gamma}_3 \to \widetilde{\Gamma}_1,
\qquad
\widetilde{\Gamma}_4 \to \widetilde{\Gamma}_2.
\ee

Writing $\Tr (\rho^{T_2}_A)^n = \widetilde{T}^F_n/2^{n-1}$, 
and performing the replacements in the formulas for $2\leqslant n \leqslant 5$ given in Eqs. 
(\ref{T2tilde}), (\ref{T3tilde}), (\ref{T4tilde}) and (\ref{T5tilde}), we find
\bea
\widetilde{T}^F_2 
&=&
2 \,\{\widetilde{\Gamma}_1 ,\widetilde{\Gamma}_2 \},
\\
\widetilde{T}^F_3 
&=&
-2\,{\{ \widetilde{\Gamma}_1^3 \}}
+6\,\{ \widetilde{\Gamma}_1^2 , \widetilde{\Gamma}_2 \},
\\
\widetilde{T}^F_4 
&=&
-4\,\{ \widetilde{\Gamma}_1^4 \}
+ 4\,\{ \widetilde{\Gamma}_1 , \widetilde{\Gamma}_2 , \widetilde{\Gamma}_1 , \widetilde{\Gamma}_2 \}
+ 8\,\{ \widetilde{\Gamma}_1^2 , \widetilde{\Gamma}_2^2 \},
\\
\label{Ttilde5 ff}
\widetilde{T}^F_5
&=&
-4\,\TrProdNormTilde{5} 
-20 \,\TrProdTilde{1}[4][2]
+20 \,\TrProdTilde{1}[3][2][2]
+20\,\TrProdTilde{1}[2][2][][1][][2].
\eea
Notice that the final expressions are very compact compared to the much more cumbersome spin counterparts.

\begin{figure}
\includegraphics[width=0.95\textwidth]{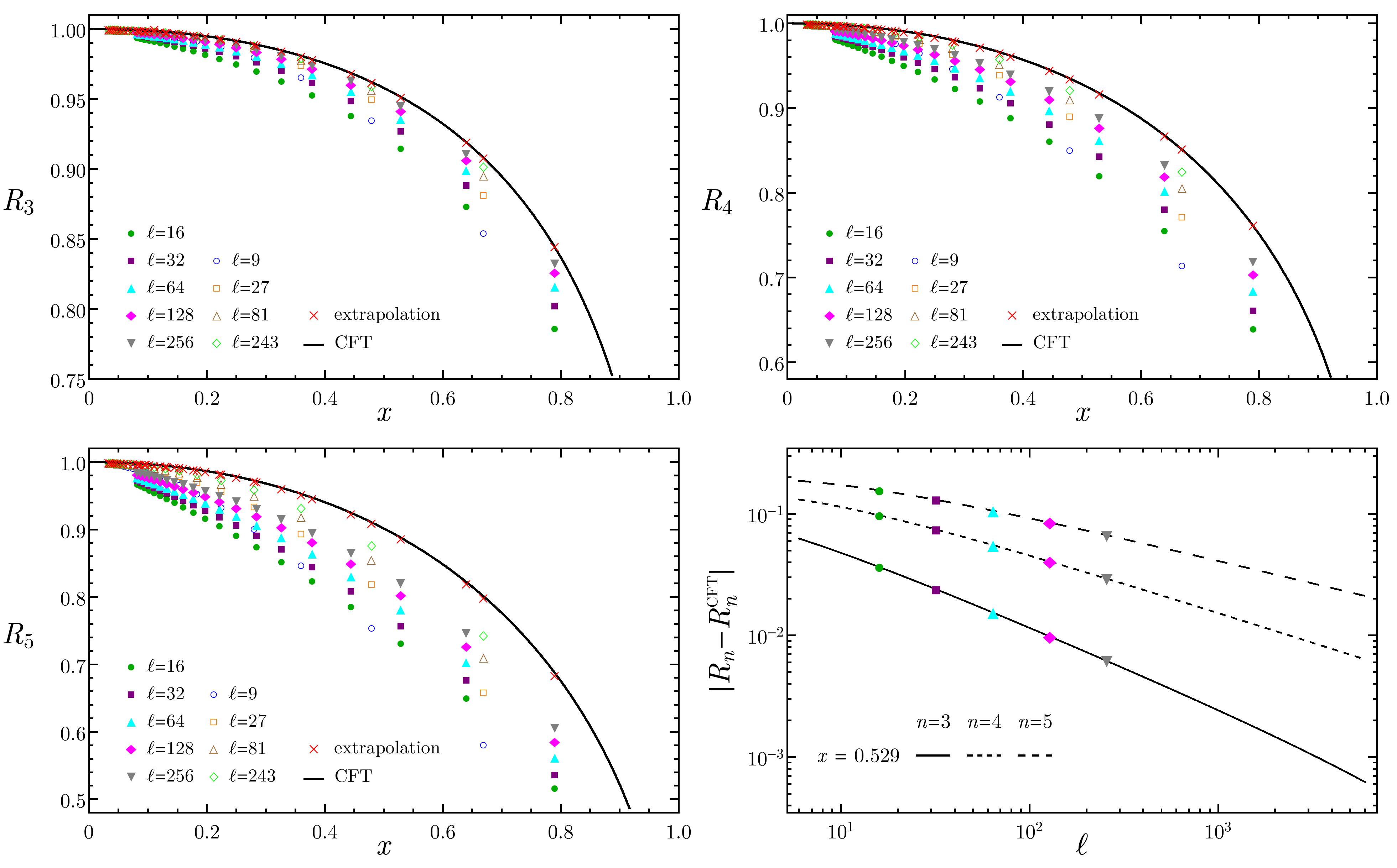}
\caption{The ratio $R_n$ between the integer moments of $\rho_A$ and $\rho_A^{T_2}$ for two 
disjoint intervals of length $\ell$ at distance $r$ for the tight-binding model at half-filling. 
We report the results for $n=3,4,5$ as function of the four-point ratio $x$ for various values of $\ell$ (and correspondingly of $r$).
For large $\ell$, the data approach the CFT predictions (solid lines).
The extrapolations to $\ell\to\infty$ --done using the scaling form (\ref{ansatz2R2})-- are shown as crosses and they perfectly agree
with the CFT curves for $n=3,4,5$.
The last panel shows explicitly the extrapolating functions for one value of $x$ and $n=3,4,5$.
} 
\label{fig:FF}
\end{figure}

\subsection{Numerical Results}

We are now going to evaluate numerically the moments of the reduced density matrix and its partial transpose. 
We can study the problem for arbitrary values of $h$ and $\gamma$ entering in Eq. (\ref{HXYF}), but in the following 
we focus on the most physically relevant fermionic system with $h=\gamma=0$, i.e. 
the tight binding model 
\be
H=\frac12 \sum_{i=1}^L \left[c^\dagger_i c_{i+1}+ c^\dagger_{i+1} c_{i} \right],
\ee
at half filling ($k_F=\pi/2$). 
In the scaling limit, the tight binding model is described by a CFT with $c=1$.

The numerical results for the ratios $R_n= \Tr (\rho^{T_2}_A)^n/(\Tr \rho_A^n)$ are reported in 
Fig. \ref{fig:FF} as function of the four-point ratio $x$ for different $\ell$ and
for $n=3,4,5$ (we checked that $R_2=1$ identically, as it should). 
We also derived  asymptotic CFT predictions for the fermionic moments of the partial transpose, 
but their derivation is too cumbersome and beyond the goals of this manuscript. 
We will report the derivation in a forthcoming publication \cite{coser-to} and we limit 
here to give the final results for $n=2,3,4,5$.
In order to have manageable formulas we introduce the shorts 
\be
\bigg[\begin{array}{c}
 2 { \boldsymbol \varepsilon}\\
 2 \boldsymbol{\delta}
 \end{array}\bigg]_\tau\,
 =
\frac{
\big| \Theta[\boldsymbol{e}]\big({\bf 0}| \tau(x)\big)\big|^2}{
\big| \Theta\big({\bf 0}| \tau(x)\big)\big|^2},
\ee
where $\Theta$ is the Riemann Theta function defined in Appendix \ref{app1}. 
In terms of the $\Theta$ function the fermionic ratios $R_n$ are given by \cite{coser-to}
\bea
& & \fl
\frac{2 R_2 }{(1-x)^{4\Delta_2}} 
 \;\longrightarrow
2 \, \bigg[\begin{array}{c}   0 \\ 1   \end{array}\bigg]_{\tilde{\tau}} ,
\\
& & \fl
\frac{4 R_3 }{(1-x)^{4\Delta_3}} 
 \;\longrightarrow
- \,2 + 6\, \bigg[\begin{array}{cc}  0 & 0 \\ 0 & 1  \end{array}\bigg]_{\tilde{\tau}},
\\
& & \fl
\frac{8 R_4 }{(1-x)^{4\Delta_4}} 
 \;\longrightarrow
- \,4 + 8\, \bigg[\begin{array}{ccc}  0 & 0 & 0 \\ 0 & 1 & 0  \end{array}\bigg]_{\tilde{\tau}} 
    + 4\, \bigg[\begin{array}{ccc}  0 & 0 & 0 \\ 1 & 1 & 1  \end{array}\bigg]_{\tilde{\tau}},
\\
& & \fl
\frac{16 R_5 }{(1-x)^{4\Delta_5}} 
 \;\longrightarrow
- \,4 + 20\, \bigg[\begin{array}{cccc}  0 & 0 & 0 & 0 \\ 0 & 0 & 1 & 0  \end{array}\bigg]_{\tilde{\tau}} 
    + 20\, \bigg[\begin{array}{cccc}  0 & 0 & 0 & 0 \\ 0 & 1 & 1 & 1  \end{array}\bigg]_{\tilde{\tau}}
    - 20\, \bigg[\begin{array}{cccc}  0 & 0 & 0 & 0 \\ 0 & 0 & 0 & 1  \end{array}\bigg]_{\tilde{\tau}},
\eea
where the matrix $\tilde{\tau}$ has been defined in Eq. (\ref{tilde tau def}) and the exponent $\Delta_n$ in App. \ref{app1}.

It is evident from Fig. \ref{fig:FF} that the lattice numerical results approach the CFT predictions 
depicted as solid lines for all $n$. 
As in the spin case, we can perform a careful finite $\ell$ analysis to take into account corrections 
to the scaling. The leading correction is expected to be of the form $\ell^{-2/n}$ and subleading 
ones to be integer powers of the leading one. 
The finite $\ell$ ansatz is then given by Eq. (\ref{ansatz2R2}) and again to have an accurate description of the data we 
keep a number of fitting parameters which make stable the extrapolation at $\ell\to\infty$.
The results of this extrapolation procedure for $n=3,4,5$ are explicitly reported in Fig. \ref{fig:FF}.
The agreement of the extrapolations with the CFT predictions is excellent also for $n=5$ as a difference 
compared to the spin counterpart.


\section{Conclusions}
\label{concl}

We have shown that the partial transpose of the reduced density matrix of two disjoint spin blocks in the XY spin chain 
can be written as a linear combination of four Gaussian fermionic operators, fully specified by their 
correlation matrices (denoted as $\widetilde\Gamma_i$, $i=1,2,3,4$ in the text) which have been explicitly calculated in terms 
of the correlation matrix of the subsystem formed by the two blocks joined with the finite part between them. 
This construction allows to calculate the moments of the partial transpose in generic configurations of the spin chain. 
In this manuscript we focused on the ground state of Ising and XY chain, but the approach is more general 
and can be used for arbitrary excited states, thermal density matrices, non-equilibrium situations etc. 

The obtained representations of the moments of the partial transpose allow us to study in an exact manner infinite chains 
and very large subsystems, drastically reducing the systematic errors in the approach to the scaling limit.
We found that for the ground state of the critical models the moments of the partial transpose agree (with high accuracy) with 
the recent CFT predictions after the corrections to the scaling are properly taken into account. 
We also studied numerically the integer powers of the partial transpose in the fermionic degrees of freedom 
(described in Ref. \cite{ez-15}, but not numerically studied). 
Even in this case we find that the moments agree perfectly with the CFT predictions in \cite{coser-to}.

The main open problem left by this manuscript for two disjoint blocks of a spin chain is whether it is somehow 
possible to obtain the negativity from the correlation matrix (the problem is also present for the fermionic degrees 
of freedom \cite{ez-15}). 
A similar problem is also open for the entanglement entropy since integer moments are obtained in a similar fashion \cite{fc-10}, 
but one has no access to the spectrum of the reduced density matrix and hence to the entanglement entropy.
From the practical point of view, it has been recently shown that if one knows a relative large number of integer moments,
rational interpolations  provide  accurate estimates of the analytic continuations \cite{ahjk-14,dct-15}
(and hence of entanglement entropy and negativity). 
However, a deeper understanding of these analytic continuations would be highly desirable.

\section*{Acknowledgments}

We are grateful to Maurizio Fagotti for many discussions on the subject of this paper.
ET is grateful to Horacio Casini for discussions. 
PC and ET thank GGI and the organisers of the workshop {\it Holographic Methods for Strongly Coupled Systems} 
for hospitality during part of this work.
PC and ET have been supported by the ERC under  Starting Grant  279391 EDEQS.

\begin{appendices}
\section{CFT results for entanglement entropy and negativity of two disjoint intervals}
\label{app1}

The moments of the reduced density matrix of two disjoint intervals for CFTs have been studied in a series of 
manuscripts \cite{ch-05,cg-08,fps-08,cct-09,c-10,headrick,cct-11,atc-11,rg-12,f-12,cz-13,hlr-13,ctt-14,RT}. 
These results have been derived using earlier findings for the partition functions of CFTs on Riemann surfaces 
with non vanishing genus \cite{bosonization higher genus}.
In this appendix we review the main results (especially from Refs. \cite{cct-neg-long,cct-09,cct-11}) which are useful 
for the comparison with numerical results. 
We mention that some universal results are also known in higher dimensions both from field theory \cite{cft-high-dims}
and holography \cite{RT,hol,rr-14}.

From global conformal invariance, we know that  $\Tr \rho_A^n$ for two disjoint intervals admits the general scaling form 
(choosing, without loss of generality, the endpoints of the intervals in the order $u_1<v_1<u_2<v_2$): 
\be
\label{tr rhoAn cft}
\Tr \rho_A^n 
\,=\,
  c_n^2 \left( \frac{(u_2-u_1)(v_2-v_1)}{(v_1-u_1)(v_2-u_2)(v_2-u_1)(u_2-v_1)} \right)^{2\Delta_n}  
  \mathcal{F}_{n}(x),
\ee
where $\Delta_n = {c}(n-1/n)/12$, being $c$ the central charge.
The variable $x$ is the four-point ratio
\be
\label{x def}
x= \frac{(u_1-v_1)(u_2-v_2)}{(u_1-u_2)(v_1-v_2)}.
\ee
Given the order of the points we have $0\leq x\leq 1$.
The prefactor $c_n$ is non-universal, but can be exactly fixed from the exact calculation of the entanglement entropy of one interval.

The difficult task of CFT is to have an exact representation for the universal function ${\cal F}_n(x)$ 
normalised so that ${\cal F}_n(0)=1$.
This universal function has been analytically derived for the compactified boson  (with central change $c=1$)  \cite{fps-08,cct-09} 
and
for the Ising CFT  (with  $c=1/2$) \cite{cct-11,atc-11}, as well as for other conformal theories which however are not of interest for this paper. 
Concerning the compactified boson, we are only interested in the value of the compactfication ratio 
corresponding to the scaling limit of the XX spin chain which is the so called Dirac point. 
For the Ising CFT and at the Dirac point
(which describe respectively the scaling limit of the critical Ising chain and the critical $XX$ model), the function ${\cal F}_{n}(x)$ reads \cite{cct-11}
\begin{equation}
\label{Fn cft def}
\mathcal{F}_{n}^{\rm Ising}(x)
= 
\frac{ \sum_{\boldsymbol{e}}
\big| \Theta[\boldsymbol{e}]\big(\tau(x)\big)\big|}{2^{n-1}\,\big| \Theta\big(\tau(x)\big)\big|},
\qquad
\mathcal{F}_{n}^{\rm Dirac}(x)
= 
\frac{ \sum_{\boldsymbol{e}}
\big| \Theta[\boldsymbol{e}]\big(\tau(x)\big)\big|^2}{2^{n-1}\,\big| \Theta\big(\tau(x)\big)\big|^2},
\end{equation}
where $\Theta[\boldsymbol{e}](\Omega) $ is the Riemann theta function, which is defined as follows  \cite{theta books}
\be\fl
\label{riemann theta def}
\Theta[\boldsymbol{e}](\Omega) 
\equiv
\sum _{\boldsymbol{m}\,\in\,\mathbb{Z}^{n-1}}
e^{
\,{\rm i} \pi
(\boldsymbol{m}+\boldsymbol{\varepsilon})^{\rm t}
\cdot\Omega\cdot(\boldsymbol{m}+\boldsymbol{\varepsilon})
+2\pi {\rm i} \,
(\boldsymbol{m}+\boldsymbol{\varepsilon})^{\rm t}\cdot \boldsymbol{\delta} 
},
\qquad
[ \boldsymbol{e}  ]
\equiv 
\bigg[\begin{array}{c}
 \boldsymbol{\varepsilon}\\
 \boldsymbol{\delta}
 \end{array}\bigg]
\equiv
\bigg[\begin{array}{c}
\varepsilon_1, \dots, \varepsilon_{n-1} \\
\delta_1, \dots, \delta_{n-1} 
\end{array}\bigg], 
\ee
being $\Omega$ a $(n-1) \times (n-1)$ symmetric complex matrix with positive imaginary part 
and $\boldsymbol{e}$ is the characteristic of the Riemann theta function, which is defined by a pair of $n-1$ 
dimensional vectors made by $\varepsilon_i, \delta_i \in \{0,1/2\}$.
In Eq. (\ref{riemann theta def}) we have to sum over all the characteristics $\boldsymbol{e}$.
The elements of the matrix $\tau(x)$ in Eq. (\ref{Fn cft def}) read \cite{cct-09}
\be\fl 
\label{tau2 matrix}
\tau(x)_{rs} = 
\textrm{i}\,\frac{2}{n} \sum_{k=1}^{n-1} \sin(\pi k/n) 
\, \frac{_2F_1(k/n,1-k/n;1;1-x)}{_2F_1(k/n,1-k/n;1;x)}\,
\cos\left [2\pi \frac{k}{n}(r-s)\right],
\ee
where $x\in (0,1)$ and $_2F_1$ is the hypergeometric function.

Also the moments of the partial transpose correspond to a four-point function of twist fields in which two 
of them have been interchanged \cite{cct-neg-letter}.
Consequently also these moments admit the universal scaling form 
\be
\label{tr rhoAn T2 cft}
\Tr \big(\rho_A^{T_2}\big)^n 
\,=\,
  c_n^2 \left( \frac{(u_2-u_1)(v_2-v_1)}{(v_1-u_1)(v_2-u_2)(v_2-u_1)(u_2-v_1)} \right)^{2\Delta_n}  
  \mathcal{G}_{n}(x),
\ee
with $c_n$ the same non-universal constant appearing in Eq. (\ref{tr rhoAn cft}) and 
$\mathcal{G}_{n}(x)$ a new universal scaling function. 
Exploiting the fact that the above moments correspond to the exchange of two twist fields, it has been 
shown that ${\cal G}_n(x)$ and ${\cal F}_n(x)$ are related as \cite{cct-neg-letter,cct-neg-long}
\be
{\cal G}_n(x) = (1-x)^{4\Delta_n} \mathcal{F}_{n}\left(\frac{x}{x-1}\right),
\ee
but some care is needed to take the analytic continuation of the function ${\cal F}_n(y)$ to 
negative argument $y$ (see for details \cite{cct-neg-long}).
This result is equivalent to say that $\Tr (\rho_A^{T_2})^n $ is given by Eq. (\ref{Fn cft def}) in which the 
period matrix $\tau(x)$ is replaced by
\be
 \tau(x)\to \tilde \tau(x)= \tau\left(\frac{x}{x-1}\right). 
 \label{tilde tau def}
\ee

Thus, the CFT prediction for ratio in Eq. (\ref{R_n def}) is 
\be
R_n^{\rm CFT}(x) = 
(1-x)^{4\Delta_n} \,\frac{\mathcal{F}_{n}\big(x/(x-1)\big)}{\mathcal{F}_{n}(x)},
\label{RnCFT}
\ee
in which the universal constants $c_n$ as well as the dimensional part of the traces 
canceled out leaving a universal scale invariant quantity. 
Thus, in order to study this quantity we do not need an a priori knowledge of the constants $c_n$
which anyhow are known both for the XX \cite{jk-04} and the Ising \cite{ij-08,ccd-09} spin chains. 
It is worth recalling that the analytic continuation to non-integer $n$ of the ratios (\ref{RnCFT}) for two intervals 
are not yet known even for the simpler cases and consequently also the negativity is eluding an 
analytic description.

%

\end{appendices}

\end{document}